\documentclass[prb,twocolumn,superscriptaddress,showpacs]{revtex4}

\usepackage[english]{babel}
\usepackage[applemac]{inputenc}
\usepackage[T1]{fontenc}
\usepackage{ae}
\usepackage{graphicx} %inlcude immagini di qualunque formato
\usepackage{amsfonts} %nuovi comandi per i font matematici
\usepackage{amssymb} %fornisce simboli addizionali
\usepackage{amsmath} %fornisce nuovi ambienti quali split, allign
\usepackage{latexsym} %altri simboli
\usepackage{psfrag} %gestione delle figure
\usepackage{float} %migliora la gestione delle immagini
\usepackage{color} %permette l'uso di colori in latex
\usepackage{subfigure} %permette di affiancare pi� figure o tabelle ognuna con la sua didascalia
\usepackage{natbib}
\usepackage[colorlinks=true,linkcolor=blue,urlcolor=blue,citecolor=blue]{hyperref}

\newcommand{\beq}{\begin{equation}}
\newcommand{\eeq}{\end{equation}}
\newcommand{\nag}{{\phantom{\dagger}}}

\def\ket#1{\mathinner{|{#1}\rangle}}

\providecommand{\abs}[1]{\lvert#1\rvert}  

\DeclareMathOperator{\Tr}{\mbox{Tr}}

\begin{document}

\title{Exploring dynamical phase transitions and prethermalization with quantum noise of excitations}

\author{Pietro Smacchia}
\affiliation{SISSA, International School for Advanced Studies, via Bonomea 265, 34136 Trieste, Italy}
\affiliation{Department of Physics and Astronomy, Rutgers University, Piscataway, New Jersey 08854, USA}
\author{Michael Knap}
\affiliation{Department of Physics, Harvard University, Cambridge MA 02138, USA}
\affiliation{ITAMP, Harvard-Smithsonian Center for Astrophysics, Cambridge, MA 02138, USA}
\author{Eugene Demler}
\affiliation{Department of Physics, Harvard University, Cambridge MA 02138, USA}
\author{Alessandro Silva}
\affiliation{SISSA, International School for Advanced Studies, via Bonomea 265, 34136 Trieste, Italy}
\affiliation{Abdus Salam ICTP, Strada Costiera 11, 34100 Trieste, Italy}

\begin{abstract}

Dynamical phase transitions can occur in isolated quantum systems that are brought out of equilibrium by  sudden parameter changes. We discuss the characterization of such dynamical phase transitions based on the statistics of produced excitations. We consider both the O(N) model in the large N limit and a spin model with long range interactions and show that the dynamical criticality of their prethermal steady-states manifests most dramatically not in the average number of excitations but in their higher moments. We argue that the growth of defect fluctuations carries unique signatures of the dynamical criticality, irrespective of the precise details of the model. Our theoretical results should be relevant to quantum quench experiments with ultracold bosonic atoms in optical lattices.

\end{abstract}

\pacs{05.30.Jp, 03.75.Kk,05.40.-a}

\maketitle

\section{Introduction}

\begin{figure}
\begin{center}
\includegraphics[width=\columnwidth]{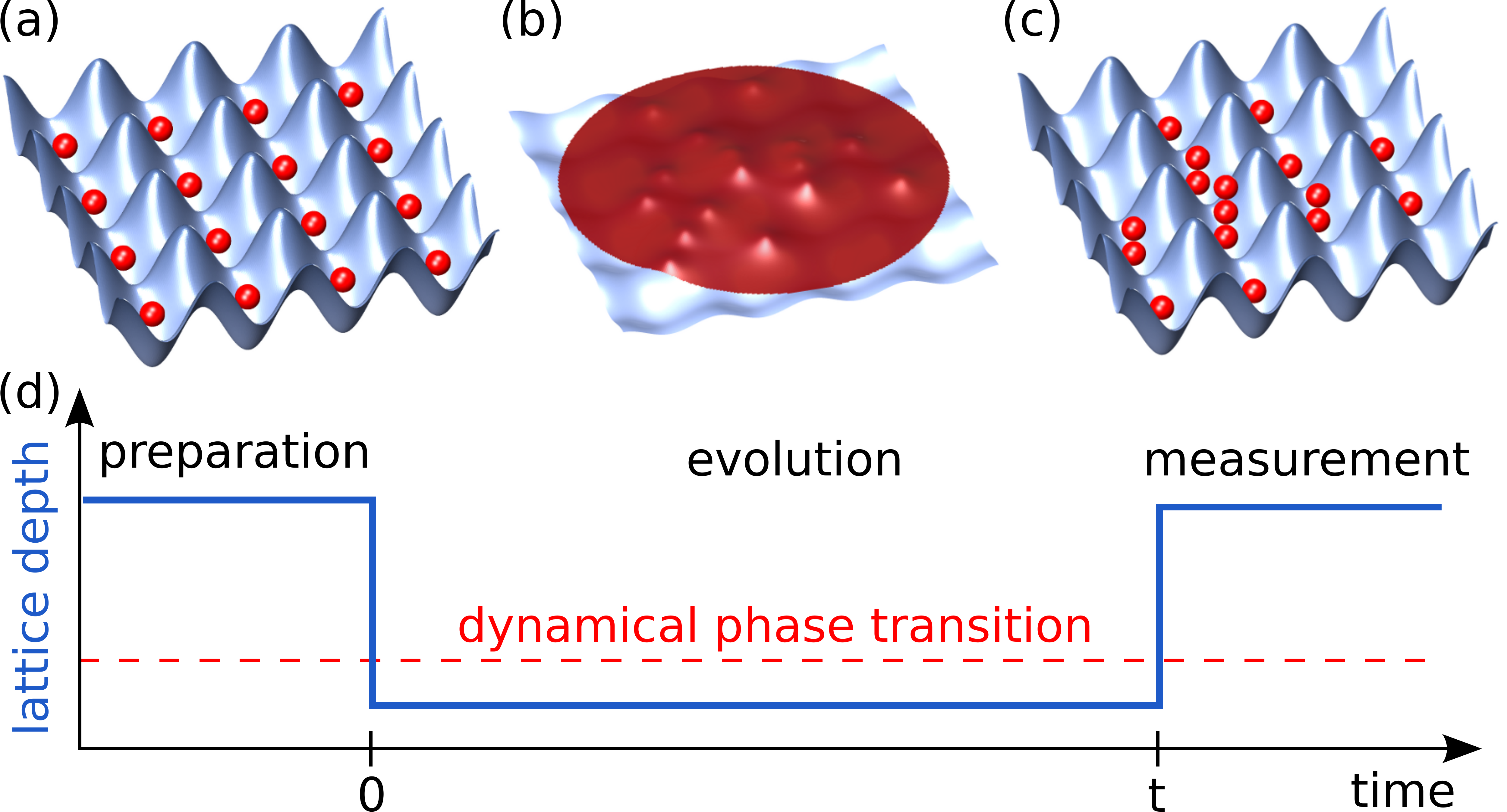}
\caption{Schematic of the double quantum quench protocol. The system is prepared in the disordered (Mott) phase in a deep optical lattice (a). By reducing the lattice depth, the system is quenched to the ordered (superfluid) phase and evolves in time (b). Finally, by rapidly ramping up the optical lattice, the dynamics is frozen and the defect density is measured (c). In (d) the lattice depth over time is shown.}
\label{fig:protocol}
\end{center}
\end{figure}

The dynamics of isolated quantum many-body systems is a subject of interest in many areas of physics involving cold atomic gases,\cite{bloch_review} solid state pump and probe experiments,\cite{Fausti2011} quantum optics,\cite{Lee2013} heavy ions collisions and cosmology. A particularly intriguing question in this context is the possible emergence and detection of new dynamical critical phenomena appearing in the steady or quasi-steady states of these systems. In this work, we will discuss possible experimental consequences of dynamical phase transitions occurring after
an abrupt change of one of the parameters of an isolated quantum system (a \it quantum quench\rm). Long times after the quantum quench a many-body system is expected to either thermalize~\cite{srednicki_94,rigol_08} or in the presence of 
integrability to relax to the Generalized Gibbs Ensemble (GGE).\cite{Jaynes1957,rigol_07} However, even when thermalization occurs, its dynamics can be highly non-trivial requiring a two step process through a prethermal state;\cite{Berges2004,moeckel_08} a phenomenon expected both in low dimensional systems close to integrability\cite{Kollar2011,Marcuzzi2013,essler_2014, nessi_2014} and in high dimensional systems close to the mean field limit. These intermediate states as well as the GGE have the intriguing feature of supporting non-thermal behavior~\cite{Kitagawa2011,Rossini2013} and in certain instances genuine dynamical critical effects, i.e. critical phenomena 
in the steady state attained after the quench.\cite{eckstein_2009,schiro_2010,sciolla_2010, sciolla_2011,Gambassi2011a,sciolla_2013} Examples of prethermalization and dynamical critical behavior 
were first observed 
in the dynamics of the Hubbard model,\cite{eckstein_2009,schiro_2010} in a variety of mean field models,\cite{sciolla_2010, sciolla_2011} and field theories\cite{Gambassi2011a} such as 
the three dimensional quantum $O(N)$ model in the infinite $N$ limit.\cite{sciolla_2013,Chandran2013}
However, the nature of these transitions and how to distinguish them from thermal critical phenomena both theoretically and, most importantly, experimentally is elusive thus far.

\begin{figure}
\begin{center}
\includegraphics[width=\columnwidth]{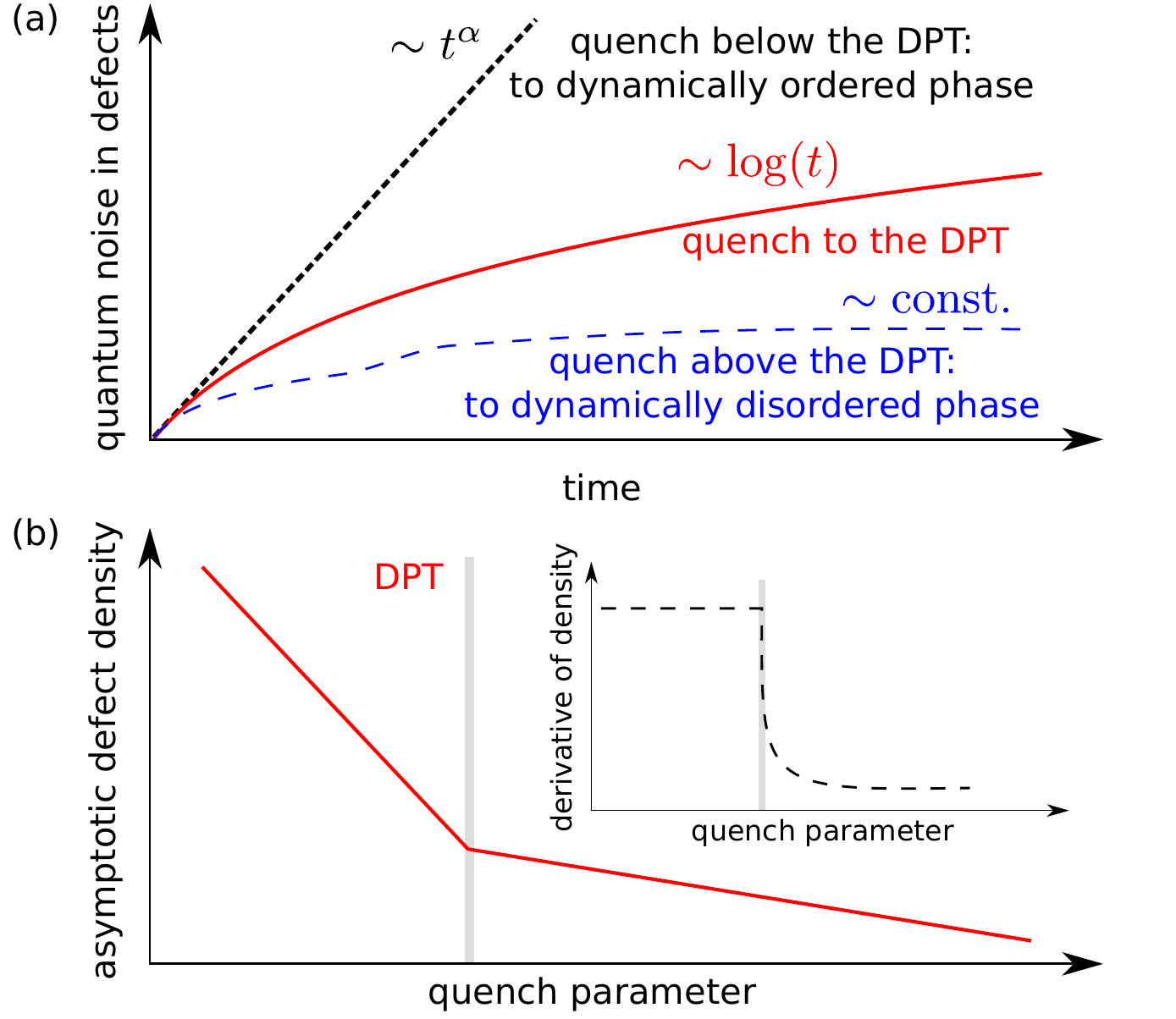}
\caption{(a) The quantum noise, i.e. fluctuations, in the number of defects, shows qualitatively different behavior for quantum quenches above, to, and below the dynamical phase transition (DPT). While it saturates for quenches to the dynamically disordered phase (above the transition), blue dashed line, it grows logarithmically for quenches to transition, red solid line, and as a power law for quenches to the dynamically ordered phase (below the transition), black dotted line. In contrast, the number of excitations in the steady state shows a non-analytic behavior as a function of the quench parameter at the dynamical critical point (b). }
\label{fig:schematicDPT}
\end{center}
\end{figure}

In this work, we address these issues and discuss a characterization of dynamical critical phenomena in bosonic systems based on the full statistics of excitations generated in a double quantum quench (see Fig.~\ref{fig:protocol}). 
In particular, we will argue that dynamical phase transitions can be detected by 
studying \it qualitatively \rm how the fluctuations in the number of excitations grow in time. Alternatively, they can be characterized by studying \it quantitatively \rm the non-analytic behavior of the stationary number of excitations (or higher moments) as a function of the quench parameter (see Fig.~\ref{fig:schematicDPT}). Experiments of this type are feasible with cold atomic gases,
where high resolution optical imaging techniques give a unique opportunity to study the dynamics of cold atoms in optical lattices  with single site resolution,\cite{bakr_2009, bakr_2010,sherson_2010} as demonstrated by recent measurements of the defects produced by ramping a system across a quantum critical point,\cite{bakr_2010} the first direct measure of a string order parameter,\cite{endres_2011} the detection of light-cone spreading of correlations,\cite{cheneau_2012} and the study of the dynamics of a mobile spin impurity.\cite{fukuhara_2013}

In order to corroborate our claims we will work out in detail the example of the quantum $O(N)$ model in the large $N$ limit, which in equilibrium and for $N=2$ is in the same universality class as the Bose-Hubbard model. The quantum $O(N)$ model is known to display a genuine dynamical phase transition for large $N$ and dimensions $d>2$. Furthermore, we discuss the infinite range Ising model {to demonstrate} that our claims are insensitive to the precise choice of the model.

A characterization in terms of traditional critical exponents would suggest that the dynamical transition of the $O(N)$ model is of the same universality as the corresponding thermal phase transition~\cite{Chandran2013}. In contrast, the full statistics of defects clearly differs from the thermal case and characterizes the dynamical criticality: while the number fluctuations of defects saturate in time for quenches above the dynamical critical point (i.e., quenches to the dynamically disordered phase), they grow indefinitely for quenches to or below the dynamical critical point (i.e., to the dynamically ordered phase), see Fig.~\ref{fig:schematicDPT}a. 
Furthermore, observables that saturate as a function of time display singularities at the dynamical transition, as shown in Fig.~\ref{fig:schematicDPT}b. 
Finite $N$ corrections are expected to eventually lead to a saturation of this indefinite growth and to a smearing of the kinks at times $\propto N$. The fate of the dynamical transition when relaxing the large $N$ constraint is beyond the scope of this work. However, our results could help to experimentally identify  dynamical criticality in systems for which theoretical results are currently not available.

The rest of the paper is organized as follows. In Sec. \ref{sec:dynamical_transition} we discuss the appearance of a dynamical phase transition in the quantum $O(N)$ model and its characterization in terms of traditional critical exponents. In Sec. \ref{sec:excitations} we compute the statistics of excitations generated in such a model by a double quench protocol as represented in Fig. \ref{fig:protocol} and show the emergence of the different qualitatively behaviors described above. In Sec. \ref{sec:ising} we discuss the case of the infinite range Ising model, showing that critical signatures in the statics of the excitations are not a unique feature of the $O(N)$ model. In Sec. \ref{sec:1d_bosehubbard} we discuss the one dimensional Bose-Hubbard model, where no prethermal behavior is expected, showing that in this case the statistics of excitations unveil the corresponding dynamical crossover diagram.  Section \ref{sec:conclusions} summarizes and discusses the results.

\section{The $O(N)$ model}
\label{sec:ONModel}
\subsection{Dynamical phase transition}
\label{sec:dynamical_transition}

The quantum $O(N)$ model consists of an $N$ component real scalar field in $d$ spatial dimensions with quartic interaction, whose Hamiltonian reads
\beq
H= \int \frac{d^d x}{2} \left[\left(\vec{\Pi}\right)^2+\left(\vec{\nabla}\vec{\phi}\right)^2+r_0 \left(\vec{\phi}\right)^2+\frac{\lambda}{12 N}\left(\vec{\phi}\right)^4 \right],
\eeq
where $[\phi_i(\vec{x}),\Pi_j(\vec{x}')]=i \delta^d(\vec{x}-\vec{x}') \delta_{ij}$, with $i$ and $j$ denoting different components. Below we will consider the $N \rightarrow \infty$ limit (see Ref.~\onlinecite{moshe_2003} for an introduction) where the model is soluble. In the disordered phase, where $\langle \phi \rangle =0$, it can be described by a quadratic theory with an effective mass parameter
\beq
r=r_0+\frac{\lambda}{12} \int_k \, \frac{1}{ \sqrt{\abs{\vec{k}}^2+r}},
\label{eq:effective_mass0}
\eeq
where from now on $\int_k=\int^\Lambda \frac{d^d k}{(2 \pi)^d}$, and $\Lambda$ is the ultraviolet cutoff. The equilibrium critical point is identified by the condition $r=0$, giving $r_{0,c}=-\frac{\lambda}{12} \int_k \frac{1}{\abs{\vec{k}}}$, which is finite for $d>1$. From Eq. (\ref{eq:effective_mass0}) it is also possible to compute the critical exponent $\nu$, since $\xi^{-1} \sim \sqrt{r}$, obtaining 
$\nu=1/2$ (mean field) for $d\geq 3$, and $\nu= \frac{1}{d-1}$ for $1<d<3$.

Let us now imagine to prepare the system in the ground state for $r_0=r_{0,i}$, corresponding to an effective mass $r_i$ and perform a quench to $r_{0,f}$. Numerical evidence for a dynamical transition following a quench of $r_0$ starting within the ordered phase has been found in this model in $d=3$.\cite{sciolla_2013} Below we will instead consider quenches starting in the disordered phase,\cite{Chandran2013} look for the dynamical critical point, i.e., the point at which the asymptotic effective mass vanishes, and calculate how the full statistics of excitations evolves in time. 

The dynamics of the system can also be described by an effective quadratic model, but the self-consistently determined effective mass becomes time-dependent and is given by
\beq
r(t)=r_{0,f}+\frac{\lambda}{6} \int_k \langle \phi_{\vec{k}}(t)  \phi_{-\vec{k}}(t)\rangle,
\label{eq:r_evolution}
\eeq
where $\phi$ represents one of the components of the field. From now on we will focus on a single component due to their inherent symmetry. Expanding the field in terms of the operators $a_{\vec{k}}$ and $a^\dagger_{\vec{k}}$
\beq
\phi_{\vec{k}}(t)=f_{\vec{k}}(t)a_{\vec{k}}+f^\star_{\vec{k}}(t)a^\dagger_{-\vec{k}},
\eeq
which diagonalize the initial Hamiltonian, i.e. $H_0=\int_k (\abs{\vec{k}}^2+r)^{1/2}\, \left(a^\dagger_{\vec{k}} a_{\vec{k}}+1/2\right)$, 
and imposing the Heisenberg equation of motions, we find that the functions $f_{\vec{k}}(t)$ have to satisfy the equation
\begin{subequations} 
\begin{align}
\frac{d^2f_{\vec{k}}(t)}{dt^2} &+\left(\abs{\vec{k}}^2+r(t) \right) \!\!f_{\vec{k}}(t)=0,\\
r(t)&=r_{0,f}+\frac{\lambda}{6} \int_k \abs{f_{\vec{k}}(t)}^2
\end{align}
 \label{eq:mainevolution}
\end{subequations}
with initial conditions $f_k(0)=\frac{1}{\sqrt{2 \omega_{k,i}}}$, $\dot{f}_k(0)=-i \sqrt{\frac{\omega_{k,i}}{2}}$, $\omega_{k,i}=\sqrt{\abs{\vec{k}}^2+r_i}$, which are fixed by the requirement that $a_{\vec{k}}$ and $a^\dagger_{\vec{k}}$ diagonalize the initial Hamiltonian.

The numerical integration \footnote{All the numerical results shown in the manuscript were obtained by setting $\Lambda=\pi$, however, the convergence of the results has been checked by varying the cutoff.} of Eqs. (\ref{eq:mainevolution}) shows that $r(t)$ always relaxes to a stationary value different from the equilibrium as a result of the fact that the distribution of quasiparticles after the quench remains non-thermal, see appendix \ref{sec:free_stationary}. To predict this stationary value we make the ansatz that the stationary part of the equal time two-body Green function is the same as in a free theory ($\lambda=0$) with initial parameter $r_i$ and final parameter $r^\star$ to be self-consistently determined.\cite{SC10} Following this route, we obtain 
\beq
r^\star = r_{0,f}+\frac{\lambda}{24} \int_k \frac{2 \abs{\vec{k}}^2+r_i+r^\star}{(\abs{\vec{k}}^2+r^\star) \sqrt{\abs{\vec{k}}^2+r_i}}.
\label{eq:mainrstar}
\eeq 
A comparison with the exact integration of Eq. (\ref{eq:mainevolution}) shows that this ansatz gives the correct asymptotic value as long as $r_{0,f} $ is above or at the dynamical transition, identified by the condition $r^\star=0$. When $r_{0,f}<r_{0,f}^c$ it predicts a negative value, while the numerical solution for the asymptotic value is always zero. Using Eq.~(\ref{eq:mainrstar}) one obtains
\beq
r^c_{0,f}=-\frac{\lambda}{24} \int_k \frac{2 \abs{\vec{k}}^2+r_i}{\abs{\vec{k}}^2 \sqrt{\abs{\vec{k}}^2+r_i}}.
\label{eq:critical_point}
\eeq
We notice that $r_{0,f}^c$ is finite for $d>2$, which thus is the lower critical dimension of the transition. Furthermore, $r_{0,f}^c$ is always less than $r_{0,c}$, i.e., always within the zero temperature ordered phase. 

From Eq. (\ref{eq:mainrstar}) it is also possible to derive the behavior of the asymptotic mass $r^\star$ for small deviations of  $r_{0,f}$ from the dynamical critical point, $\delta r_{0,f}=r_{0,f}-r_{0,f}^c$. For $\delta r_{0,f}>0$ we then have
\beq
r^\star=\delta r_{0,f}- \frac{\lambda}{6} r^\star \int_k \frac{\sqrt{\abs{\vec{k}}^2+r_i}}{4 \abs{\vec{k}}^2(\abs{\vec{k}}^2 +r^\star) }.
\eeq
For $d>4$ the integral is convergent in the limit $r^\star \rightarrow 0$, so that $r^\star \sim \delta r_{0,f}$, while for $ 2 <d <4$ the integral is the dominant term implying $r^\star \sim \left(\delta r_{0,f} \right)^{2/(d-2)}$. This translates to the behavior of the correlation length in the stationary state $\xi^\star$, since $(\xi^\star)^{-1} \sim \sqrt{r^\star}$. Defining the exponent $\nu^\star$ as $(\xi^\star)^{-1} \sim \left(\delta r_{0,f}\right)^{\nu^\star}$, we thus have $\nu^\star=\frac{1}{d-2}$ for $2<d<4$ and $\nu^\star=1/2$ for $d \geq 4$, with $d=4$ playing the role of an upper critical dimension.

Apparently the critical properties described above are similar to that of the finite temperature transition,\cite{Chandran2013} i.e., critical dimensions and exponents are obtained by a shift up by one dimension as compared to the corresponding quantum phase transition. However, we will now show that, contrary to the thermal case, the dynamical transition leaves strong signatures on the statistics of excitations produced in the quantum quench.\\

\subsection{Statistics of excitations and signatures of critical behaviour}
\label{sec:excitations}

\begin{figure}
\begin{center}
\includegraphics[width=\columnwidth]{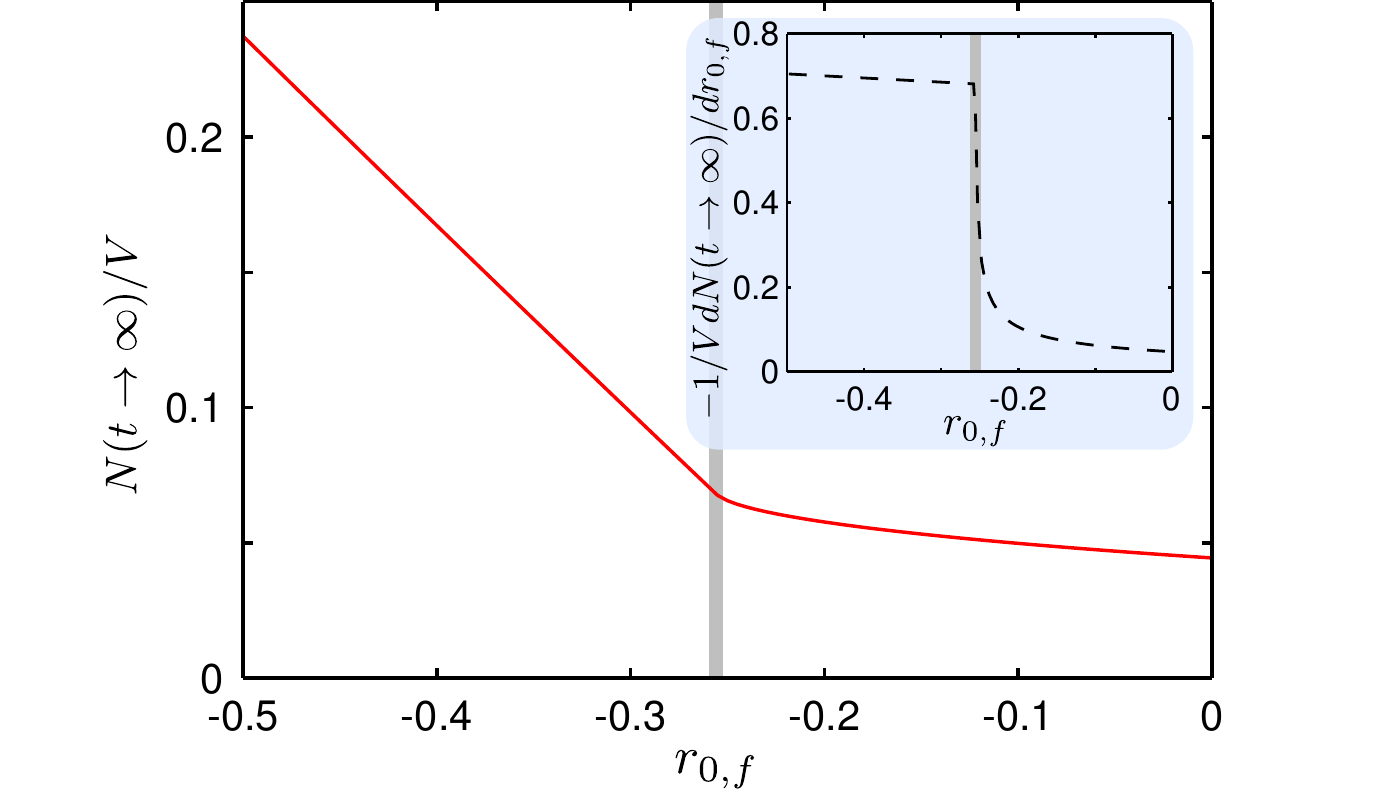}
\caption{ (Color online) Saturation value of the defect density $N(t\to \infty)/V$, $V=L^d$, red solid line, for quenches from $r_{0,i}=5$ to $r_{0,f}$ and $\lambda=10$. $N(t\to \infty)/V$ is non-analytic at the dynamical phase transition, indicated by the thick gray line. Inset: The first derivative of the asymptotic defect density $-\frac{1}{V} \frac{dN(t\to \infty)}{dr_{0,f}}$ exhibits a pronounced kink at the transition.}
\label{fig:nSat}
\end{center}
\end{figure}

Let us now imagine starting in the disordered phase and performing a first quench of $r_0$ at or close to, the dynamical critical point (see Fig. \ref{fig:protocol}). We then let the system evolve for a time $t$ and finally return to $r_{0,i}$ in order to count the number of excitations generated, and observable described by the operator $\hat{N}=\int_k a^\dagger_{\vec{k}} a_{\vec{k}}$ \footnote{We note that the statistics of this operator cannot be simply reduced to the structure factor $\langle \int_k \phi_{\vec{k}} \phi_{-\vec{k}} \rangle $ (or 
its powers), since the operators $\phi_{\vec{k}}$ and $a_{\vec{k}}$ do not commute in general.}. This is definitely a 
fluctuating quantity characterized by a probability distribution $P(N,t)$, which equivalently can be described in terms of the moment generating function $G(s,t)=\langle e^{-s \hat{N}} \rangle_t$. For the $O(N)$ model in the large $N$ limit, this quantity can be computed exactly. Indeed, since the theory is effectively quadratic and the different $k$-modes interacts only through the renormalization of the mass $r(t)$, we obtain $G(s,t)=\prod_{\vec{k}} G_{\vec{k}}(s,t)$ with $G_{\vec{k}}(s,t)$ representing the generating function for a single mode.

In order to compute $G_{\vec{k}}(s,t)$ we first express the time evolved state $\ket{\psi(t)}_{\vec{k}}$ as a function of $a_{\vec{k}}$ and $a^\dagger_{\vec{k}}$. The starting point is the expansion of the time evolved field $\phi_{\vec{k}}(t)$ in the same basis, which can be translated from Heisenberg to Schr\"{o}dinger picture by writing
\begin{subequations}
\beq
\phi_{\vec{k}}(0)=f_{\vec{k}}(t) \tilde{a}_{\vec{k}}(t)+f_{\vec{k}}^\star(t) \tilde{a}^\dagger_{-\vec{k}}(t),
\eeq
\beq
\Pi_{\vec{k}}(0)=\dot{f}_{\vec{k}}(t) \tilde{a}_{\vec{k}}(t)+\dot{f}_{\vec{k}}^\star(t) \tilde{a}^\dagger_{-\vec{k}}(t).
\eeq
\label{eq:tilde_relation}
\end{subequations}
The operators $\tilde{a}_{\vec{k}}$ and $\tilde{a}^\dagger_{-\vec{k}}$ are defined by the relation $\tilde{a}_{\vec{k}}(t) \ket{\psi(t)}=0$, i.e. they annihilate the time evolved state. At the same time, we know that 
\begin{subequations}
\beq 
\phi_{\vec{k}}(0)=	\frac{1}{\sqrt{2 \omega_{k,i}}} \left(a_{\vec{k}}+a^\dagger_{-\vec{k}}\right),
\eeq
\beq
\Pi_{\vec{k}}(0)=	i\sqrt{\frac{\omega_{k,i}}{2}} \left(a^\dagger_{-\vec{k}}-a_{\vec{k}}\right).
\eeq
\label{eq:zero_relations}
\end{subequations}
By inverting Eq. (\ref{eq:tilde_relation}), taking into account that $f_{\vec{k}}(t) \dot{f}_{\vec{k}}^\star(t)-\dot{f}_{\vec{k}}(t)f_{\vec{k}}^\star(t)=i$, and inserting the result into Eq. (\ref{eq:zero_relations}), one obtains
\beq
\tilde{a}_{\vec{k}}(t)=\alpha_{\vec{k}}^\star(t) a_{\vec{k}}-\beta^\star_{\vec{k}}(t) a^\dagger_{-\vec{k}},
\label{eq:bogoliubov}
\eeq
with
\begin{subequations}
 \beq
\alpha_{\vec{k}}(t)=f_{\vec{k}}(t) \sqrt{\frac{\omega_{k,i}}{2}}+i \frac{\dot{f}_{\vec{k}}(t)}{\sqrt{2 \omega_{k,i}}},
\eeq
\beq
\beta_{\vec{k}}(t)=f_{\vec{k}}(t) \sqrt{\frac{\omega_{k,i}}{2}}-i \frac{\dot{f}_{\vec{k}}(t)}{\sqrt{2 \omega_{k,i}}}.
\label{eq:beta_def}
\eeq
\end{subequations}
From Eq. (\ref{eq:bogoliubov}) and the requirement that $\tilde{a}_{\vec{k}}(t)$ annihilates the time evolved state, one finally finds
\beq
\ket{\psi(t)}_k=\frac{1}{\sqrt{\abs{\alpha_{\vec{k}}(t)}}} \exp \left(\frac{\beta_{\vec{k}}^\star(t)}{2 \alpha_{\vec{k}}^\star(t)} a^\dagger_{\vec{k}} a^\dagger_{-\vec{k}} \right) \ket{0},
\eeq
with $a_{\vec{k}}\ket{0}=0$.

Having the expression of the state in terms of $a_{\vec{k}}$ and $a^\dagger_{\vec{k}}$, the computation of $G_k(s,t)$ can be straightforwardly done, for example using coherent states. 
We finally get $G(s,t)=\exp(-L^d f(s,t))$ with
\beq
f(s,t)=\frac{1}{2}\int_k \log \left[1+ \rho_k(t) \left(1-e^{-2 s} \right) \right],
\label{eq:char_func}
\eeq
defined for $s> -\bar{s}=\frac{1}{2} {\rm sup}_k \log \frac{\rho_k(t)}{1+\rho_k(t)}$. Here, $L$ is the linear size of the system and 
\beq
\rho_k(t)=\abs{\beta_{\vec{k}}}^2 = \abs{f_{\vec{k}}(t)}^2 \frac{\omega_{k,i}}{2}+\frac{\abs{\dot f_{\vec{k}}(t)}^2}{2 \omega_{k,i}}-1/2,
\label{eq:rho_def}
\eeq
with $k= \abs{\vec{k}}$. The function $\rho_k$, which fully determines the statistics of the excitations, is obtained from integrating Eq. (\ref{eq:mainevolution}) and represents the average number of excitations in each mode.

\begin{figure}
\begin{center}
\includegraphics[width=\columnwidth]{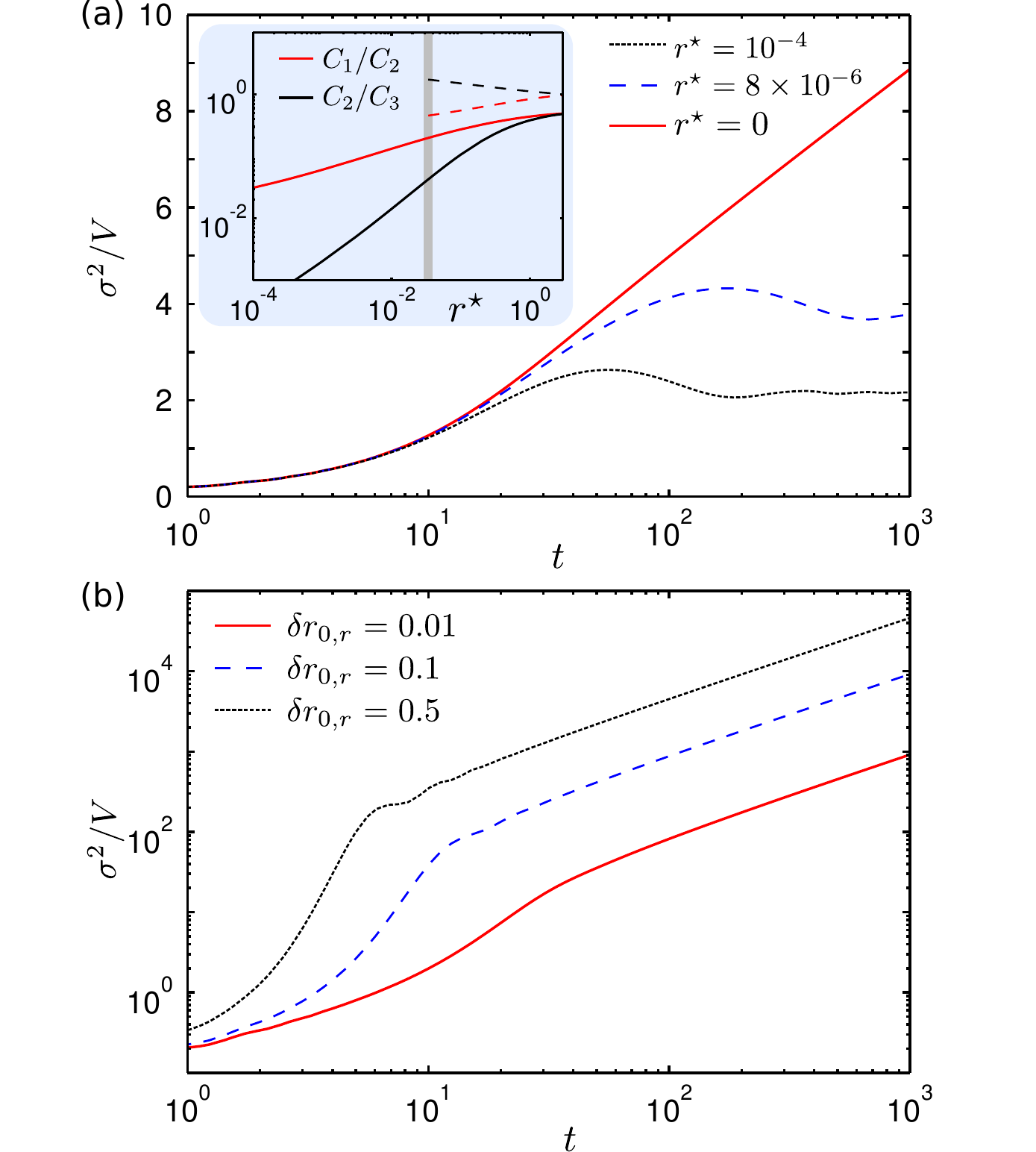}
\caption{ (Color online) (a) Variance per unit volume $\sigma^2/V$, $V=L^d$, in a log-linear scale for quenches above or at the dynamical transition, i.e., $r_{0,f}\geq r_{0,f}^c$ in $d=3$ for different values of the predicted asymptotic effective parameter $r^\star$, see Eq. (\ref{eq:mainrstar}). The inset shows ratios of different cumulants $C_n(t\to \infty)$ as a function of $r^\star$, solid lines, and compares them to the corresponding equilibrium cumulants at finite temperature, dashed lines. (b) Variance per unit volume $\sigma^2/V$ in a log-log scale for quenches below the dynamical transition, i.e., $r_{0,f}<r_{0,f}^c$ in $d=3$. $\delta r_{0,f}= r_{0,f}^c-r_{0,f}$ measures the distance from the dynamical critical point. In all plots $\lambda=10$ and $r_i=5$.}
\label{fig:cumulants_O(N)}
\end{center}
\end{figure}

Let us now characterize the dynamical critical behavior of the system by studying all the cumulants $C_n$'s of the distribution of excitations, using the formula $C_n (t)=(-1)^n \frac{\partial^n}{\partial s^n} \log G(s,t) |_{s=0}$. Below, we present the first two cumulants, i.e., the average $N(t)$ and the variance $\sigma^2(t)$, in $d=3$ and discuss their characteristic dynamics for quenches to intermediate values of the bare mass $r_{0,f}$ which are above, below, or at the dynamical critical point. For additional data see appendix \ref{sec:supplemental_results}.

First of all, it is important to notice that the time evolution of the average and of the variance are qualitatively different.
The former does not display striking features and saturates for all $r_{0,f}$ in the long time limit. However, its asymptotic value as a function of $r_{0,f}$ displays non-analytic behavior at the dynamical critical point, see Fig. \ref{fig:nSat}. We observe similar non-analyticities also in the infinite range Ising model, as discussed in section \ref{sec:ising}. In contrast, the variance exhibits three qualitatively different behaviors, as schematically introduced in Fig.~\ref{fig:schematicDPT}.
When the first quench is at the dynamical critical point, i.e. $r_{0,f}=r_{0,f}^c$, the variance per unit volume grows logarithmically in time $t$, see Fig. \ref{fig:cumulants_O(N)}a. This should be contrasted with what one would expect for a free field theory, where the variance grows linearly in $d=3$, see appendix \ref{sec:free_stationary}. A totally different behavior is observed for quenches below the dynamical critical point (to the dynamically ordered phase), i.e. $r_{0,f} < r_{0,f}^c$: in this case the variance grows as a power law $t^{\alpha}$  with $\alpha=1$ in $d=3$, 
Fig. \ref{fig:cumulants_O(N)}b. Finally, for quenches to an intermediate value of the bare mass above the dynamical transition (to the dynamically disordered phase), i.e. $r_{0,f} > r_{0,f}^c$, the variance saturates at a finite value, Fig. \ref{fig:cumulants_O(N)}a. 

The physical motivation to explore higher moments of the excitations is that they probe the small momentum modes which inevitably characterize dynamical criticality. More specifically, the statistics of the excitations and the scaling of all the cumulants for large times $t$ is fully determined by the scaling of $\rho_k(t)$ for small $k$. Indeed $\rho_k(t)$ is singular as $1/k^\gamma$ up to an infrared cutoff shrinking to zero as $1/t$. Since the $n$-th cumulant is given by a weighted sum of the integrals over $k$ of all the integer powers of $\rho_k$ up to $n$, we can infer that its asymptotic behavior in $t$ is given by 
\beq
C_n \sim \int_{1/t} dk \,k^{d-1-\gamma n} \sim t^{\gamma n-d}.
\eeq
Numerical results in $d=3$ confirm that, as expected from the behavior of the variance, $\gamma=3/2$ for quenches to the critical point and $\gamma=2$ for quenches below the critical point, while in $d=4$ we have $\gamma=2$ and $\gamma=3$ in the two respective cases.\cite{supp}

% \subsection{Finite size cumulants}

\begin{figure}[t]
\begin{center}
\includegraphics[width=0.45\textwidth]{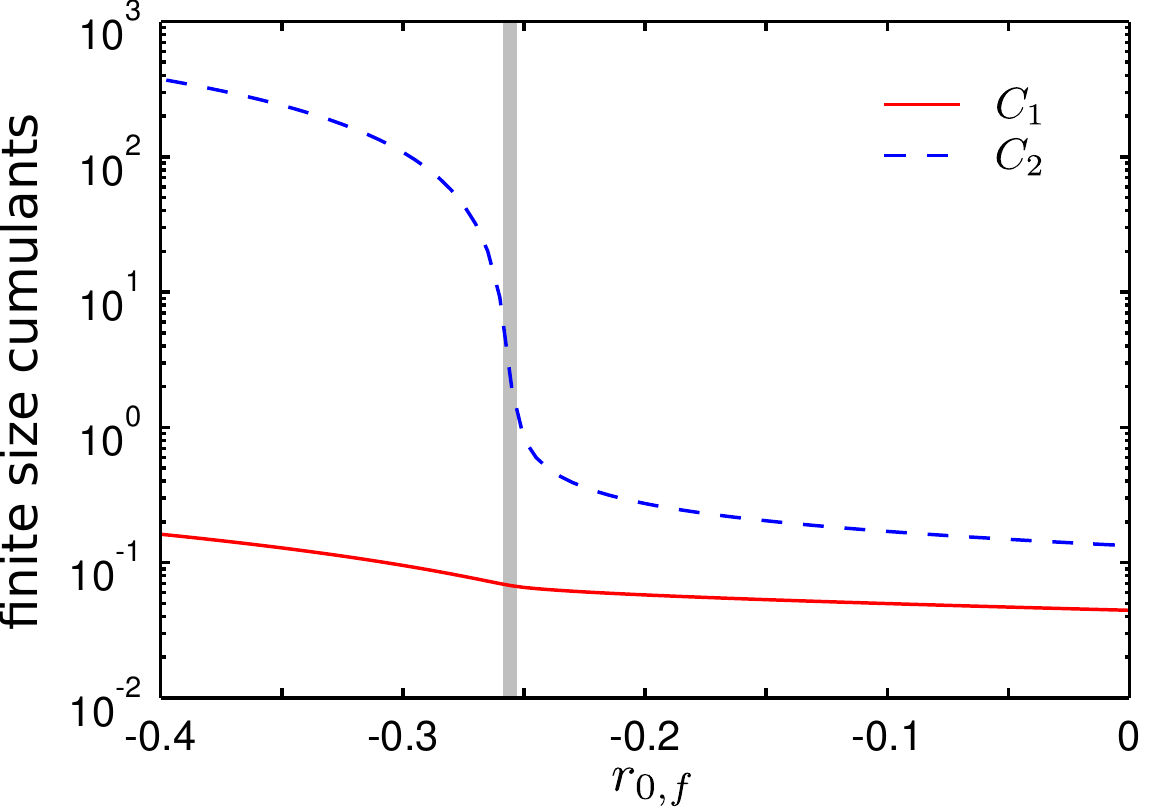}
\caption{ Long-time saturation value of the first two cumulants for a finite volume as a function of the quench parameter $r_{0,f}$. They show signatures of the dynamical phase transition, indicated by the tick gray solid line. }
\label{fig:supp3} 
\end{center}
\end{figure}

Let us now briefly discuss how the above discussed behavior of the cumulants is affected by the presence of a finite (but  still large) volume  $V=L^3$. This scenario can be described by an infrared cutoff $\sim \pi/L$ in the integrals over the momenta. In this case, the variance does not grow indefinitely but rather saturates as a function of time. Therefore, the time dependence observed before applies only to the transient. However, there are still signatures of the dynamical transition in the behavior of the saturation value as a function of the quench parameter $r_{0,f}$, as one may easily ascertain from Fig.~\ref{fig:supp3}.\\
 
\section{Infinite range Ising model}
\label{sec:ising}

As stated in the Introduction, we expect the connection between the statistics of the excitations and dynamical phase transitions not to be limited to the specific case of the $O(N)$ model in the large $N$ limit. To corroborate such a statement, we will briefly discuss here the simpler case of the infinite range Ising model, which is also known to display a dynamical phase transition.\cite{sciolla_2011} 

The Hamiltonian of the infinite range Ising model reads
\beq
\begin{split}
H_I&=-\frac{K}{2N}\sum_{i,j=1}^N \sigma^z_i \sigma^z_j-\lambda \sum_i \sigma^x_i\\
&=-\frac{K}{2N} (S^z)^2-\lambda S^x
\end{split}
\eeq
where $S^{\alpha}=\sum_i \sigma^{\alpha}_i$ and $N$ is the total number of spins.

Differently from the $O(N)$ model studied in the main text, the statistics of the excitations of this model coincides with the statistics of the magnetization, i.e., the number of spin flips along the $z$ directions. The computation can be readily done, assuming that the initial state is a coherent state (and so is its subsequent time evolution), 
\beq
|\theta,\varphi\rangle=e^{\frac{1}{2}\theta e^{i\varphi}S_{-}-\frac{1}{2} \theta e^{-i\varphi}S_{+}} |N/2\rangle.
\eeq
From this we obtain for the generating function
\beq
G(s)=\langle \theta,\varphi | e^{-s S^z} | \theta,\varphi \rangle = e^{s N/2}\left(\frac{e^{-s}+|z|^2}{1+|z|^2}\right)^N\;.
\eeq
The final step is to obtain differential equations for the parameters $\theta(t)$ and $\phi(t)$, which can be achieved starting from the Bloch equations for the spin operators, taking averages $\langle S^i S^j \rangle \simeq \langle S^i \rangle \langle S^j \rangle$, and using the parametrization $S^z=\frac{N}{2}\cos\theta,
S^x=\frac{N}{2}\sin\theta\cos\varphi,
S^y=\frac{N}{2}\sin\theta\sin\varphi$. After some lines of calculation, the final result is 
\begin{subequations}
\begin{eqnarray}
\partial_t\theta&=&\lambda \sin\varphi,\label{dynamics1}\\
\sin\theta\partial_t \varphi&=&-\frac{K}{2}\cos\theta\sin\theta+\lambda \cos\varphi\cos\theta.\label{dynamics2}
\end{eqnarray}
\end{subequations}
Solving these equations gives the time dependent statistics of the excitations. For a quantum quench starting in the ferromagnetic phase to a certain $\lambda_f$, one finds oscillatory solution for both $\theta(t)$ and $\phi(t)$, so we actually focus on the statistics of the time averaged magnetization $\overline{S^z}$.

\begin{figure}
\begin{center}
\includegraphics[width=0.46\textwidth]{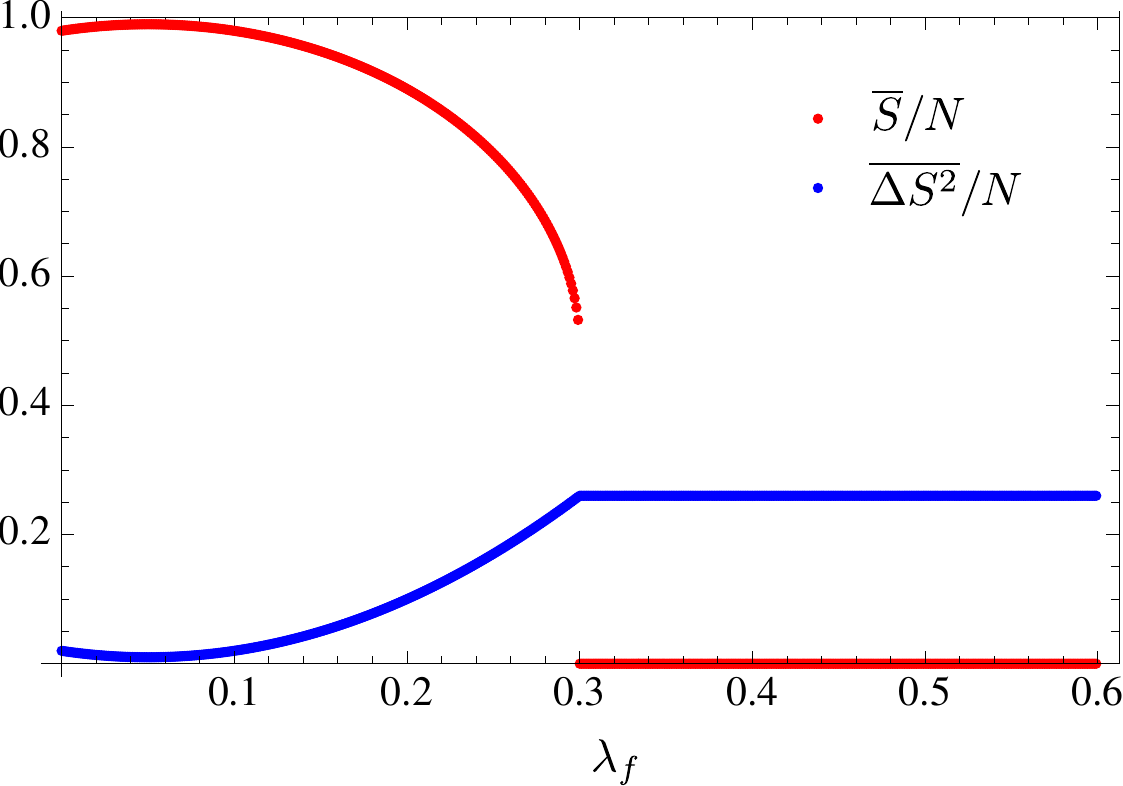}\\
\end{center}
\caption{First two cumulants of the time average statistics of the magnetization as a function of the final quench parameter $\lambda_f$ starting from the ground state at $\lambda_i=0.1$.}
\label{fig:magnetization_statistics}
\end{figure}
\begin{figure*}
\begin{center}
\includegraphics[width=0.88\textwidth]{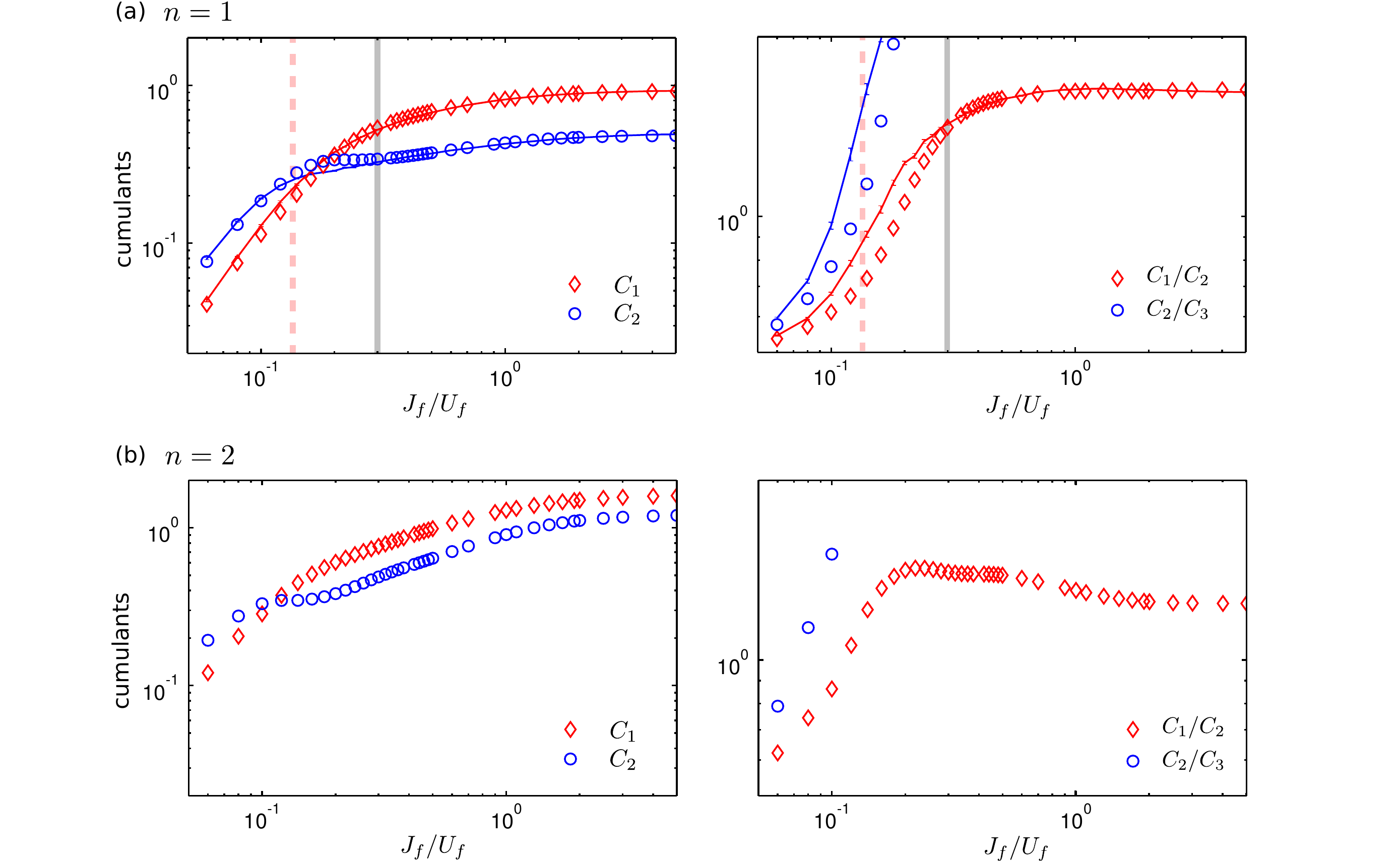}
\caption{ Statistics of the global defect density in the nonintegrable, one-dimensional Bose-Hubbard model. First and second cumulant as a function of the quench parameter $J_f/U_f$, left, and their ratios, right, at filling (a) $n=1$ and (b) $n=2$. Vertical lines indicate the equilibrium phase transition, solid gray line, and the ratio  $J_f/U_f$ at which the equilibrium gap in the Mott phase corresponds to the energy density pumped into the system, dashed line. The thin solid lines show the thermal value of the cumulants at the self-consistently determined effective temperature.}
\label{fig:supp2}
\end{center}
\end{figure*}

In Fig. \ref{fig:magnetization_statistics} we show the behavior of the first two cumulants per unit spin at long times as a function of the final quench parameter $\lambda_f$ for a fixed initial state, corresponding to the ground state at $\lambda_i=0.1$. We can see that both the cumulants signals the presence of the dynamical phase transition at $\lambda_f=0.3$ trough their non-analyticities.

We stress that, even though in the above discussed case of the Ising model measuring the statistics of the excitations and looking at the order parameter is essentially equivalent, this is not true in general, as in the case of the $O(N)$ model and of Hubbard-like models.

\section{One dimensional Bose-Hubbard model}
\label{sec:1d_bosehubbard}

The statistics of excitations may be a useful quantity to study even when no dynamical criticality is expected. In order to illustrate this, in this section we study the quench dynamics in the nonintegrable, one-dimensional Bose-Hubbard model using exact numerical techniques, where our simulations provide insights into the nonequilibrium crossover diagram taking into account the full many-body interactions. 

We introduce the Bose-Hubbard model on a lattice
\begin{equation}
 \hat{H}_\text{BH}=-J \sum_{\left\langle i,\,j \right\rangle} \left(b_i^\dag   \, b_j^\nag + \text{h.c.}\right) + \frac{U}{2} \sum_i \hat{n}_i(\hat{n}_i-1 ) -\mu \sum_i \hat{n}_i\;,
\end{equation}
where $J$ is the kinetic energy, $U$ the interaction energy, and $\mu$ the chemical potential. The boson creation and annihilation operators are $b_i^\dag$ and $b_i^\nag$, respectively, which define the density operator $\hat n_i= b^\dag_i b^\nag_i$. 

We follow the protocol of the double quench introduced in the Introduction (see Fig. \ref{fig:protocol}), by starting out deep in the disordered phase at commensurate filling with $J_i/U_i=0.01$, where the ground state $|\psi_0 \rangle$ is close to a product state. The dynamics is initialized by quenching the kinetic energy to $J_f/U_f$. Consequently the system evolves for the wait time $t$ at which the statistics of global defects
\beq
\hat D=\sum_i |\hat n_i-n|
\eeq 
is measured, where $n$ is the density of bosons. Higher cumulants can be obtained from the generating function in the usual way. In our simulations all cumulants saturate. We attribute this to the fact that the nonlinearities are fully treated in the exact simulations and therefore the unbounded growth observed in the field theory gets regularized. 

In Fig.~\ref{fig:supp2} we show the saturation value of the first and second cumulant (i.e., the average and variance, respectively) normalized by the volume of the system for various quench parameters $J_f/U_f$ and commensurate density $n=1$ and $n=2$. We find that the more energy is pumped into the system by the quantum quench, i.e., the larger the final kinetic energy $J_f/U_f$ is, the larger is the saturation value of the global defect density $D$, and its higher order statistics.

Since the Bose-Hubbard model is not integrable, it is expected to thermalize. To study this effect, we perform finite temperature simulations in which the effective temperature $T^*$ is self-consistently determined by the energy density pumped into the system by the quantum quench
\begin{equation}
\langle\psi_0 |H_\text{BH}(J_f,U_f)|\psi_0 \rangle=\frac{\Tr[H_\text{BH}(J_f,U_f)e^{-{H_\text{BH}(J_f,U_f)}/{T^\star}}]}{\Tr[e^{-{H_\text{BH}(J_f,U_f)}/{T^\star}}]},
\end{equation}
where $\ket{\psi_0}$ is the state that initializes the dynamics which is the ground state of $H_\text{BH}(J_i,U_i)$. The statistics of excitations evaluated in the thermal state are indicated by the thin solid lines in Fig.~\ref{fig:supp2}(a) and support thermalization for large values of $J_f/U_f$ already after a few inverse hopping times. At low values of $J_f/U_f \lesssim 0.3$, which marks the equilibrium phase transition, gray thick line, small deviations between the thermal and the long time average can be observed. However, we study rather small systems of $L=8$ sites and in order to make a conclusive statement in that regime a proper finite size scaling needs to be done. The thick dashed line shows the ratio of $J_f/U_f$ at which the equilibrium gap corresponds to the energy density of the quantum quench. 
Around this coupling the deviations of the global defect statistics from the thermalized state seem to be largest. Note, however, that the deviations from the thermal results are vanishingly small for all $J_f/U_f$ when we consider the statistics of local (instead of global) defects, not shown.  In the right column of Fig.~\ref{fig:supp2}(a) we show the ratio of the cumulants. The trend here is that the larger $J_f/U_f$, the larger the ratios $C_1/C_2$ and $C_2/C_3$, which is opposite to the prediction of the field theory for higher dimension, see inset of Fig.~4~(a). 

An important difference between the Bose-Hubbard model at low filling and the field theory is the following: While infinitely many particle excitations can be created in the Bose-Hubbard model on top of a certain state with commensurate filling $n$, only $n$ holes can be created locally. This has to be contrasted with the field theory which does not obviously discriminate between particle and hole excitations. Therefore, one could expect, that for nonequilibrium dynamics the agreement between field theory and the Bose-Hubbard model improves at higher filling. In Fig.~\ref{fig:supp2}(b) we thus show the saturation values when starting out at filling $n=2$. The main difference here is that the ratio $C_1/C_2$ decreases for larger $J_f/U_f$ similarly to the results obtained from the field theory in higher dimension.

\begin{figure*}[]
\includegraphics[width=\textwidth]{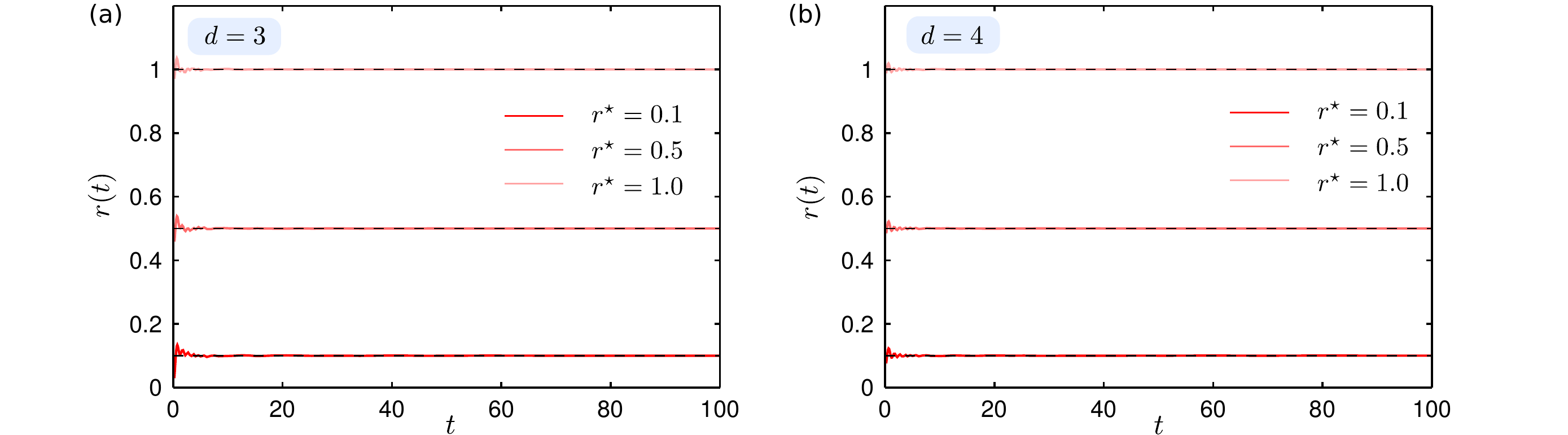}
\caption{(Color online) Comparison between $r(t)$ obtained by numerical integration of Eq.(\ref{eq:evolution}) for quenches to different $r_{0,f}>r_{0,f}^c$ (curves of different colors) and the asymptotic value predicted by Eq. (\ref{eq:rstar}) (black dashed lines) for $d=3$ (a) and $d=4$ (b).}
\label{fig:mass}
\end{figure*}

\section{Conclusions}
\label{sec:conclusions}

In conclusion, we observed that the dynamical phase transition of the quantum $O(N)$ model in the large $N$ limit leaves a strong imprint on the statistics of excitations generated in a quantum quench. We expect this phenomenon to be generic for systems where dynamical transitions are known to be present at mean field level. We corroborated this claim by also studying the infinite range Ising model in Sec. \ref{sec:ising} which displays similar behavior. 

Whether signatures of such dynamical transition can be observed in realistic systems such as the Bose-Hubbard model is an important open question. We argued that the statistics of excitations could be an experimentally accessible tool to solve this problem. Indeed,
even though the excitations will not grow indefinitely in a real experimental system, the dynamical phase transition can still leave a unique fingerprint on the statistics of excitations in the intermediate prethermal state before full thermalization occurs.
Experimental studies with ultracold atoms might therefore be able to shed light on this challenging question.

\section{Acknowledgment}

We thank G. Biroli, A. Gambassi, A. Polkovnikov for useful discussions.
The authors acknowledge support from Harvard-MIT CUA, 
ARO-MURI Quism program, ARO-MURI on Atomtronics, 
as well as the Austrian Science Fund (FWF) Project No. J 3361-N20. 

\appendix

\section{Free theory and stationary state}
\label{sec:free_stationary}

\begin{figure*}
\begin{center}
\includegraphics[width=0.98\textwidth]{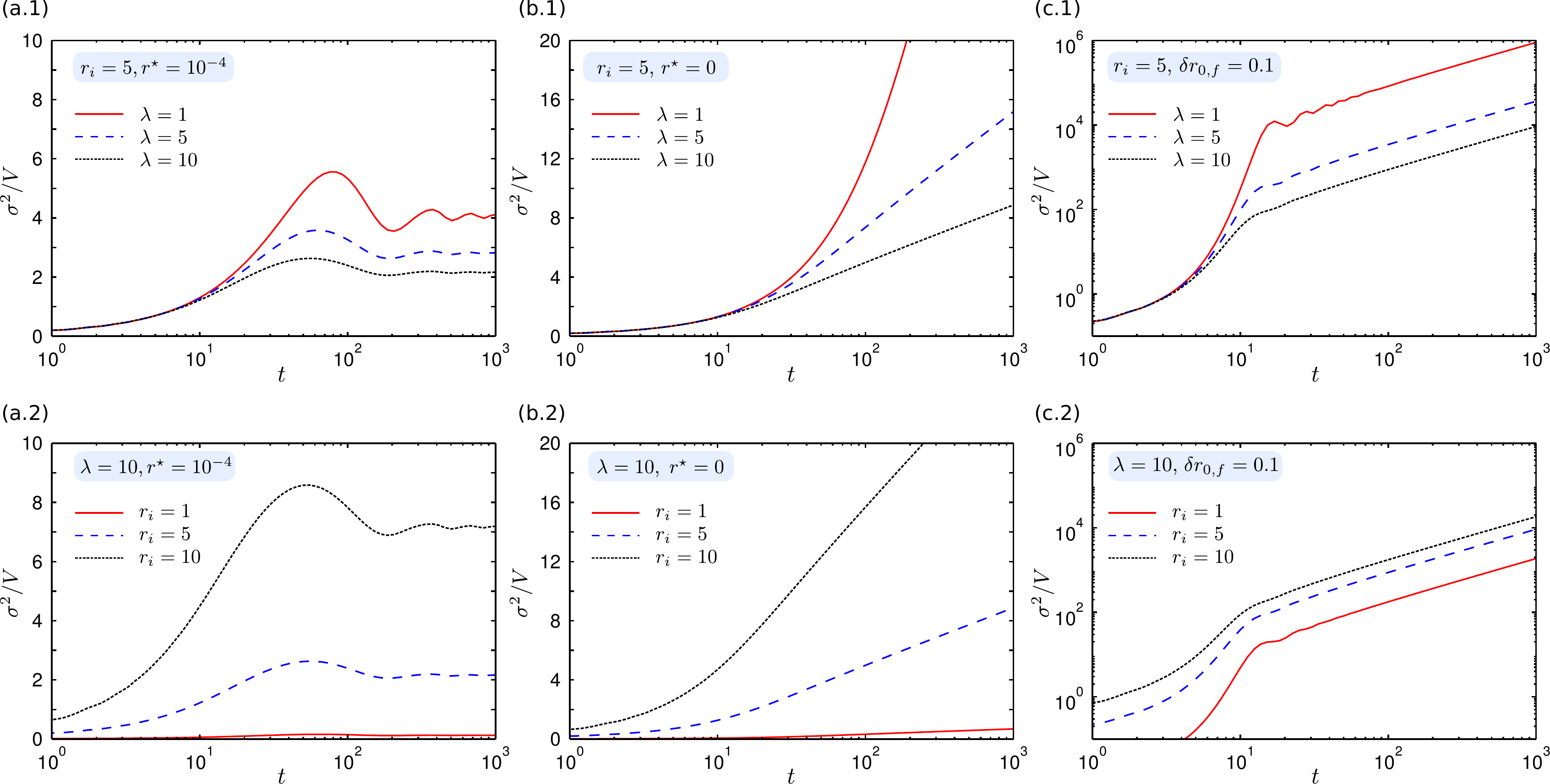}
\caption{ Fluctuations of excitations for quenches (a) above, (b) to, and (c) below the dynamic phase transition for various parameters as indicated in the figure caption and legend. }
\label{fig:supp1}
\end{center}
\end{figure*}

\begin{figure*}
\begin{center}
\includegraphics[width=0.98\textwidth]{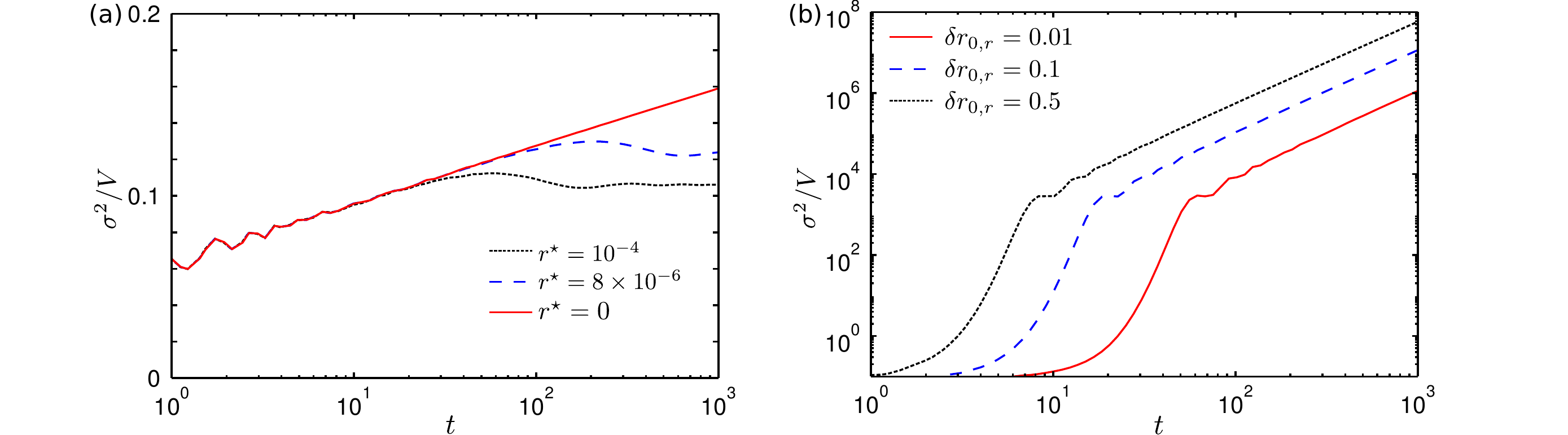}
\caption{(Color online) (a) Variance per unit volume $\sigma^2/V$ in log-linear scale for quenches above or at the dynamical transition, i.e. $r_{0,f}\geq r_{0,f}^c$ in $d=4$ for different values of the predicted asymptotic effective parameter $r^\star$, see Eq. (\ref{eq:rstar}).  (b) Variance per unit volume $\sigma^2/V$, $V=L^d$, in log-log scale for quenches below the dynamical transition, i.e. $r_{0,f}<r_{0,f}^c$ in $d=4$. $\delta r_{0,f}= r_{0,f}^c-r_{0,f}$ measures the distance from the dynamical critical point. In all plots $\lambda=10$ and $r_i=5$.}
\label{fig:supp4}
\end{center}
\end{figure*}

In this section we will consider the dynamics of the systems for a quench of $r_0$ from $r_{0,i}$ to $r_{0,f}$, when there is no quartic interaction, i.e, $\lambda=0$. The first obvious consequences are that there is no renormalization of the initial mass and also no dynamics of the mass after the quench. The equation for the functions $f_{\vec{k}}(t)$, which are the coefficients of the expansion of the field $\phi_{\vec{k}}(t)$ in the basis of the initial Hamiltonian, becomes
\beq
\frac{d^2f_{\vec{k}}(t)}{dt^2} +\left(\abs{\vec{k}}^2+r_{0,f} \right) \!\!f_{\vec{k}}(t)=0,
\label{eq:evolution}
\eeq
with initial conditions $f_k(0)=\frac{1}{\sqrt{2 \omega_{k,i}}}$, $\dot{f}_k(0)=-i \sqrt{\frac{\omega_{k,i}}{2}}$, $\omega_{k,i}=\sqrt{\abs{\vec{k}}^2+r_{0,i}}$, set by the requirement that $a_{\vec{k}}$ and $a^\dagger_{\vec{k}}$ diagonalize the initial Hamiltonian.

The solution of the previous equation is readily found to be $f_{\vec{k}}(t)=\frac{1}{\sqrt{2 
\omega_{k,i}}} \cos \left(t \sqrt{\abs{\vec{k}}^2+r_{0,f}}\right)-\frac{i}{\sqrt{\abs{\vec{k}}^2+r_{0,f}}} \sqrt{\frac{\omega_{k,i}}{2}} \sin \left(t \sqrt{\abs{\vec{k}}^2+r_{0,f}}\right)$. From this expression we can compute all the quantities of interest, including the equal time two-point correlator of the field $\langle \phi_{\vec{k}}(t)  \phi_{-\vec{k}}(t)\rangle=\abs{f_{\vec{k}}(t)}^2$:
\beq
\begin{split}
&\langle \phi_{\vec{k}}(t) \phi_{\vec{-k}}(t) \rangle = \frac{2 \abs{\vec{k}}^2+r_{0,i}+r_{0,f}}{4(\abs{\vec{k}}^2+r_{0,f}) \sqrt{\abs{\vec{k}}^2+r_{0,i}} }+\\
&\frac{r_{0,f}-r_{0,i}}{4 (\abs{\vec{k}}^2+r_{0,f}) \sqrt{\abs{\vec{k}}^2+r_{0,i}} }\cos\left(2 t\sqrt{\abs{\vec{k}}^2+r_{0,f}} \right).
\label{eq:correlation_function}
\end{split}
\eeq

Instead, in the case of the interacting theory with $\lambda \neq 0$ the time dependent effective mass is given by 
\beq
r(t)=r_{0,f}+\frac{\lambda}{6} \int_k \langle \phi_{\vec{k}}(t)  \phi_{-\vec{k}}(t)\rangle.
\label{eq:r_t}
\eeq
The numerical integration of the equation of motions shows that for large $t$ this relaxes toward a stationary value. To predict this stationary values we make the ansatz that the stationary part of the equal time Green's function $\langle \phi_{\vec{k}}(t) \phi_{-\vec{k}}(t)\rangle$ is the same as the free theory but with renormalized masses. In particular, we take Eq. (\ref{eq:correlation_function}), disregard the cosine contribution, and make the substitutions $r_{0,i} \rightarrow r_i$ and $r_{0,f} \rightarrow r^\star$, with $r^\star$ denoting the stationary value of the mass, to be self-consistently determined from Eq. (\ref{eq:r_t}). In this way we obtain the self-consistent equation for $r^\star$ written in the main text, that is
\beq
r^\star = r_{0,f}+\frac{\lambda}{24} \int_k \frac{2 \abs{\vec{k}}^2+r_i+r^\star}{(\abs{\vec{k}}^2+r^\star) \sqrt{\abs{\vec{k}}^2+r_i}}.
\label{eq:rstar}
\eeq
Fig. \ref{fig:mass} demonstrates how accurate this equation predicts the stationary value of $r(t)$ up to the dynamical critical point, identified by the condition $r^\star=0$ focusing on $d=3$ or $d=4$, but we checked Eq. (\ref{eq:rstar}) also in lower and higher dimensions.

Using the solution of the Eq. (\ref{eq:evolution}) for $\lambda=0$ and Eq. (\ref{eq:rho_def}), one can find the function $\rho_k(t)$, and thus determine the full statistics of excitations for the free case. The result of such a procedure is
\beq
\rho_k(t)=\frac{(r_{0,f}-r_{0,i})^2}{4(\abs{\vec{k}}^2+r_{0,f})(\abs{\vec{k}}^2+r_{0,i})}\sin\left( t\sqrt{\abs{\vec{k}}^2+r_{0,f}} \right)^2.
\label{eq:rho_k}
\eeq
As discussed at the end of Sec. \ref{sec:excitations}, from this expression, and in particular from its low-$k$ behavior, one can extract the behavior of all cumulants. We see that, apart from the sine which provides an infrared cutoff evolving as $1/t$, for $r_{0,f} \neq 0$, $\rho_k$ is regular at low $k$, while for $r_{0,f}=0$, which is the critical point of the free theory, $\rho_k \sim 1/k^2$. This implies $k_n \sim t^{2n -d}$, with $k_n$ denoting the $n$-th cumulant and $t^0$ corresponds to a subleading logarithmic growth.

\section{Supplemental results}
\label{sec:supplemental_results}

A systematic study of the time dependent fluctuations in $d=3$ for various parameters of our model is shown in Fig.~\ref{fig:supp1}.

In addition, in Fig~\ref{fig:supp4} the results obtained for the variance in $d=4$ are shown. The variance shows again three different qualitative behavior: saturation when $r_{0,f}$ is above the dynamical critical point, logarithmic growth when $r_{0,f}$ is at the dynamical critical point, and power law growth with exponent of two when $r_{0,f}$ is below the dynamical critical point.

\bibliography{Nonequilibriumv_2}

\end{document}